\documentclass[fleqn,twoside]{article}
\usepackage{msty}
\usepackage{a4wide,epsfig}

\readRCS
$Id: espcrc2.tex,v 1.2 2004/02/24 11:22:11 spepping Exp $
\usepackage{graphicx}
\usepackage[figuresright]{rotating}

\newcommand{\AmS}{{\protect\the\textfont2
  A\kern-.1667em\lower.5ex\hbox{M}\kern-.125emS}}

\title{$\mbox{BR}(\bar{B}\to X_s\gamma)$: First Next-to-Next-to-Leading
  Order Results}

\author{Matthias Steinhauser\address[]{II. Institut f\"ur Theoretische Physik,
        Universit\"at Hamburg,\\
          Luruper Chaussee 149, D-22761, Hamburg, Germany.} 
        \thanks{To appear in the Proceedings of the International Conference
        and Loops and Legs in Quantum Field Theory 2004, Germany, 
        25-30 April 2004.}
}
       
\runtitle{}
\runauthor{M. Steinhauser}

\begin{document}

\begin{abstract}
In this contribution the first next-to-next-to-leading order
corrections to the branching ratio
$\mbox{BR}(\bar{B}\to X_s\gamma)$ are discussed.
This includes the completion of the matching part and the fermionic
corrections to the most important operator matrix elements.
\vspace{1pc}
\end{abstract}

% typeset front matter (including abstract)
\maketitle

%%%%%%%%%%%%%%%%%%%%%%%%%%%%%%%%%%%%%%%%%%%%%%%%%%%%%%%%%%%%

\section{Motivation}

\begin{figure}[t]
  \begin{center}
    \epsfig{file=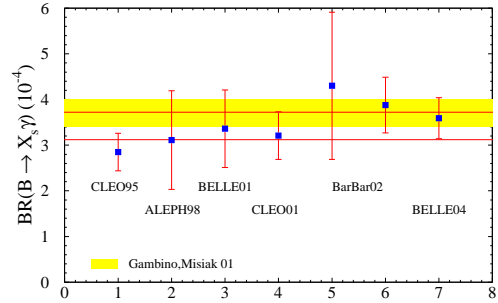,width=7.0cm}
    \vspace*{-1cm}
    \caption{\label{fig::bsgexp}Comparison of the experimental
    measurements for 
    $\mbox{BR}[\bar{B} \to X_s \gamma]$, where a cut-off on the photon
    energy $E_{\gamma} > m_b/20$ is applied,
    with the theoretical predictions given by the (yellow)
    band. The full (horizontal) lines indicate the uncertainties of the
    combined experimental results (cf. Eq.~(\ref{eq::bsg.exp})).}
  \end{center}
\vspace*{-1cm}
\end{figure}

The rare decay of a $\bar{B}$ meson into a meson containing a strange
quark and a photon is very sensitive to new physics which is connected to the
fact that in the Standard Model (SM) this process is loop-induced.
Thus a precise comparison of theoretical calculations and experimental
measurements can lead to important hints on the theory beyond the SM.
Conventionally the decay is normalized to the semileptonic $B$-meson
decay. The corresponding branching ratio has meanwhile been measured
by CLEO~\cite{Chen:2001fj}, ALEPH~\cite{Barate:1998vz}, 
BELLE~\cite{Abe:2001hk} and BABAR~\cite{Aubert:2002pd} which
is summarized in Fig.~\ref{fig::bsgexp}
and leads to the experimental world average\footnote{The CLEO
  measurement from 1995 is not included in the average.}~\cite{Jessop.2002}
\begin{eqnarray}
  \lefteqn{\mbox{BR}[\bar{B} \to X_s \gamma,~~( E_{\gamma} > m_b/20) ] 
  =}\nonumber\\&& ( 3.42 \pm 0.30 ) \times 10^{-4}
  \,.
  \label{eq::bsg.exp}
\end{eqnarray}
This result agrees very well with the SM
predictions~\cite{Gambino:2001ew,Buras:2002tp}
\begin{eqnarray}
  \lefteqn{\mbox{BR}[\bar{B} \to X_s \gamma,~~(E_{\gamma} > 1.6~{\rm GeV}) ] 
  =}\nonumber\\&& ( 3.57 \pm 0.30 ) \times 10^{-4}\,,\nonumber\\
  \lefteqn{\mbox{BR}[\bar{B} \to X_s \gamma,~~( E_{\gamma} > m_b/20) ] 
  \simeq}\nonumber\\&& 3.70 \times 10^{-4}\,.
  \label{eq::BRtheo}
\end{eqnarray}
In Fig.~\ref{fig::bsgexp} the latter is compared to the measurements.
In the near future the experimental error will reduce to approximately
5\% which reduces the current error by almost a factor of two.
Thus also the reduction of the theoretical uncertainty
in Eq.~(\ref{eq::BRtheo}) is mandatory.

At next-to-leading order (NLO)
a detailed analysis of the theoretical error has been performed in
Ref.~\cite{Gambino:2001ew}. It has been shown that the largest
uncertainty comes form the dependence on the charm quark mass which
arises for the first time at this order. Including all
next-to-next-to-leading order (NNLO) effects
one arrives at an uncertainty of approximately 7\% 
which has to be compared with
the 8\% error quoted in Eq.~(\ref{eq::BRtheo}).
Thus a complete NNLO calculation can significantly
reduce the theoretical uncertainty.

Calculations for the process $\bar{B} \to X_s \gamma$ are
conveniently performed within the framework of an effective theory 
where the scales of order $M_W$ and higher are integrated out. 
This leads to an effective Lagrangian of the form
\begin{eqnarray}
    \label{eq::lag}
  &&\!\!\!{\cal L}_{\rm eff} = {\cal L}_{\scriptscriptstyle 
    {\rm QCD} \times {\rm QED}}(u,d,s,c,b) 
  \\ \mbox{}
  &&\!\!\!+ \frac{4 G_F}{\sqrt{2}} \sum_{i,j}  
  \left[ V^*_{cs} V_{cb} C^c_i \; + \; V^*_{ts} V_{tb} C^t_i \right] Z_{ij}
    P_j\,, \nonumber
\end{eqnarray}
where $G_F$ is the Fermi constant, $V$ stands for the
Cabibbo-Kobayashi-Maskawa (CKM) matrix, $P_j$ are the operators,
$C_i^{c/t}$ are the Wilson coefficients and $Z_{ij}$ the corresponding
renormalization constants.
More details and explicit expressions can be found in~\cite{MisSte04} and
references cited therein.

In such an effective-theory
framework the calculation splits into three steps: 
(i) the matching of the full and the effective theory at the high scale, 
(ii) the evaluation of the operator matrix elements at
the low scale $m_b$, and 
(iii) the running from the high to the low scale.
At NNLO step (i) requires the evaluation of two-loop diagrams in order
to obtain the matching coefficients for the four-fermion operators
which has been performed in Ref.~\cite{Bobeth:1999mk}.
The three-loop matching coefficients $C_7$ and $C_8$ of the
dimension-five dipole operators have been 
evaluated in Ref.~\cite{MisSte04} which will be considered in more detail
in Section~\ref{sec::C7C8}.

As far as the evaluation of the NNLO operator matrix elements are
concerned up to now only the 
fermionic corrections of ${\cal O}(\alpha_s^2 n_f)$
to $P_1$, $P_2$, $P_7$ and $P_8$
are known~\cite{Bieri:2003ue}. At NLO these operators are
numerically important which is a strong motivation to
consider the numerical effects at NNLO. Furthermore, once the
corrections proportional to $n_f$ are available, it is tempting to 
apply the so-called naive non-abelianization
in order to
estimate the complete corrections of order $\alpha_s^2$.
This is based on the observation
that the lowest coefficient of the QCD $\beta$ function,
$\beta_0=11-2n_f/3$, is quite large and thus it is expected
that the replacement of $n_f$ by $-3\beta_0/2$ may lead to a good
approximation of the full order $\alpha_s^2$ corrections.
In Section~\ref{sec::ferm} we provide more details on this
calculation.
Finally, Section~\ref{sec::concl} contains our conclusions.

%%%%%%%%%%%%%%%%%%%%%%%%%%%%%%%%%%%%%%%%%%%%%%%%%%%%%%%%%%%%

\section{\label{sec::C7C8}Three-loop contribution to $C_7$ and $C_8$}

There are several possibilities to evaluate the matching coefficients.
For instance, it is possible to evaluate the Feynman diagrams
on-shell. However, we find it more convenient to follow the procedure
outlined in Ref.~\cite{Bobeth:1999mk} where all the necessary diagrams are
evaluated off-shell, after expanding them in the external momenta. 
In intermediate steps 
spurious infrared divergences appear which are 
generated by the expansion and are regulated dimensionally.
They cancel out in the matching equation, i.e. in the
difference between the full SM and the effective theory off-shell
amplitudes.

The scalar three-loop integrals are evaluated with the help of the
package {\tt MATAD} \cite{Steinhauser:2000ry} designed for calculating
vacuum diagrams.  The fact that {\tt MATAD} can deal with a single
non-vanishing mass only is not an obstacle against taking into account
the actually different masses of the $W$ boson and the top
quark. Expansions starting from $m_t=M_W$ and $m_t \gg M_W$ allow us
to accurately determine the three-loop matching conditions for the
physical values of $m_t$ and $M_W$.

\begin{figure}[t]
\begin{center}
\includegraphics[width=4cm,angle=0]{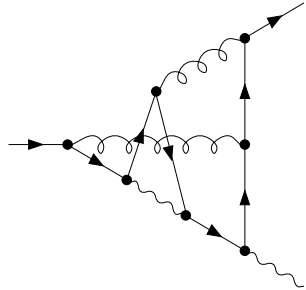}
\end{center}
\vspace*{-1cm}
%\hspace*{106mm} $s$\\[1cm]
%\hspace*{85mm} $u,c,t$\\[-1mm]
%\hspace*{64mm} $b$\\[9mm]
%\hspace*{81mm} $W$\\[2mm]
%\hspace*{105mm} $\gamma$\\[-5mm]
\begin{center}
\caption{One of the ${\cal O}(10^3)$ three-loop diagrams that 
  contributes to $b\to s\gamma$.}
\label{fig:sampleSM}
\end{center}
\vspace*{-1cm}
\end{figure}

\begin{figure*}[t]
%\vspace*{-3cm}
\begin{center}
%\hspace*{-4cm}
\includegraphics[width=7cm,angle=0]{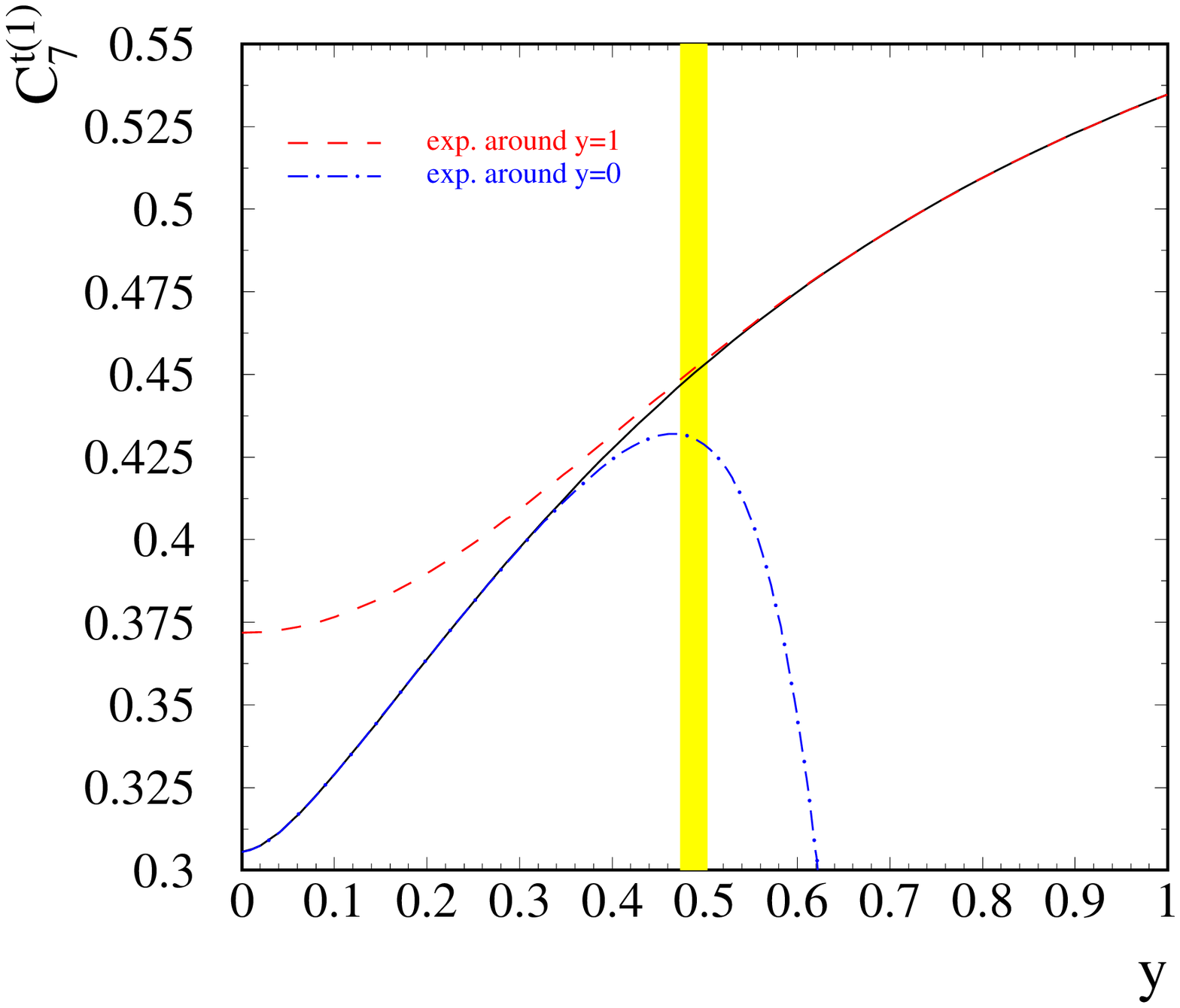}%\hspace{-7mm}
\includegraphics[width=7cm,angle=0]{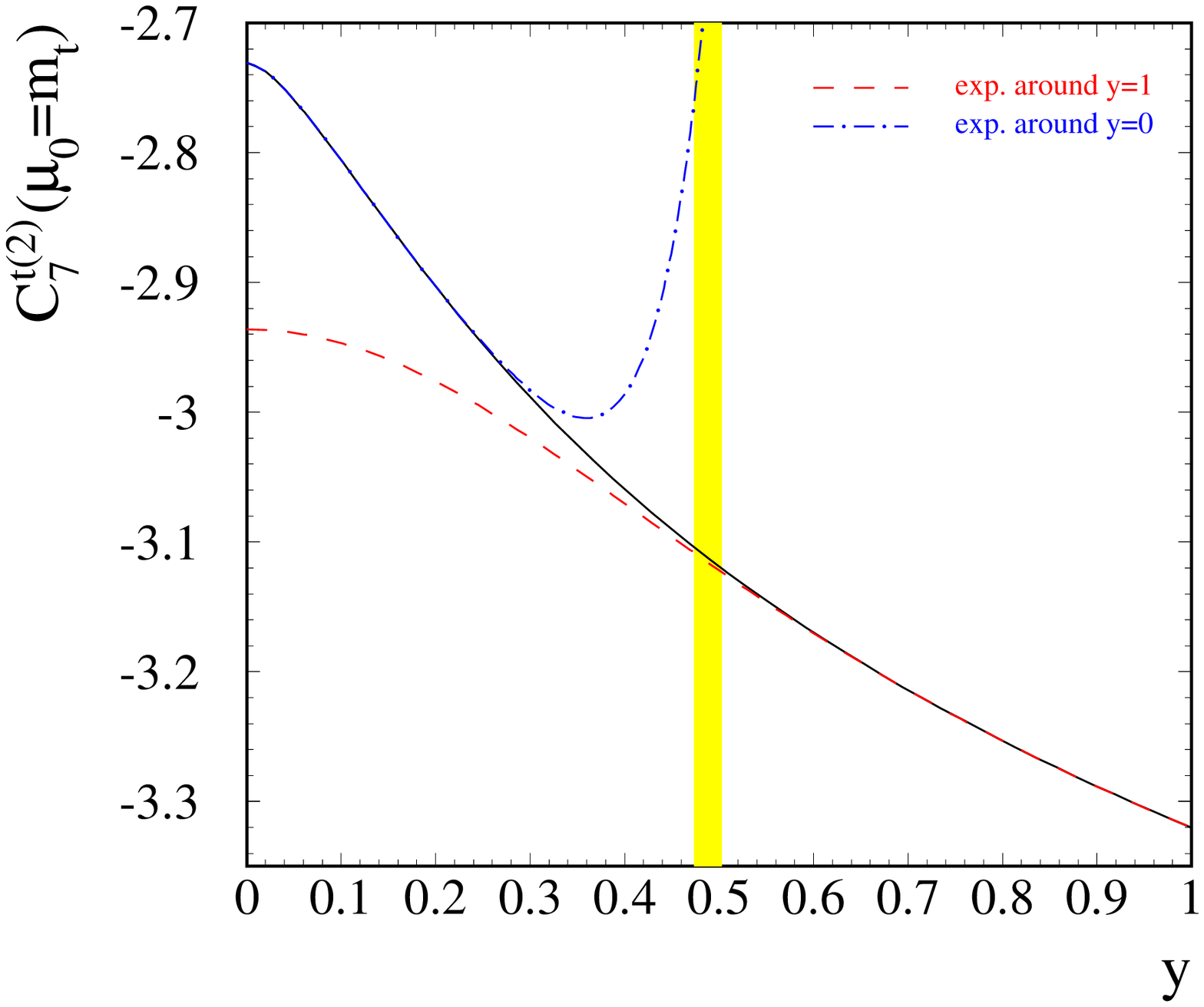}\\%\hspace{-4cm} \ \\
%\hspace*{-4cm}
\includegraphics[width=7cm,angle=0]{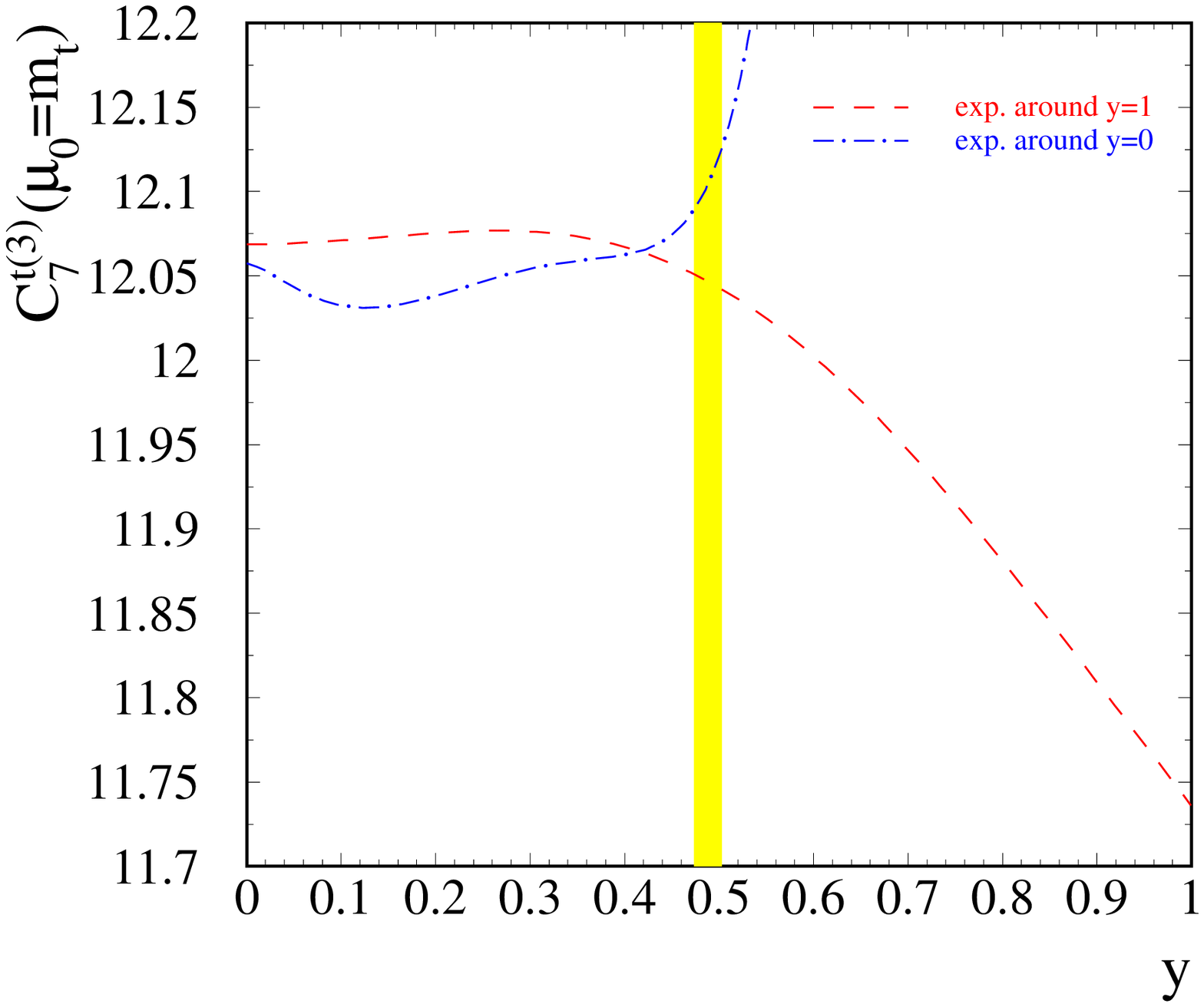}%\hspace{-7mm}
\includegraphics[width=7cm,angle=0]{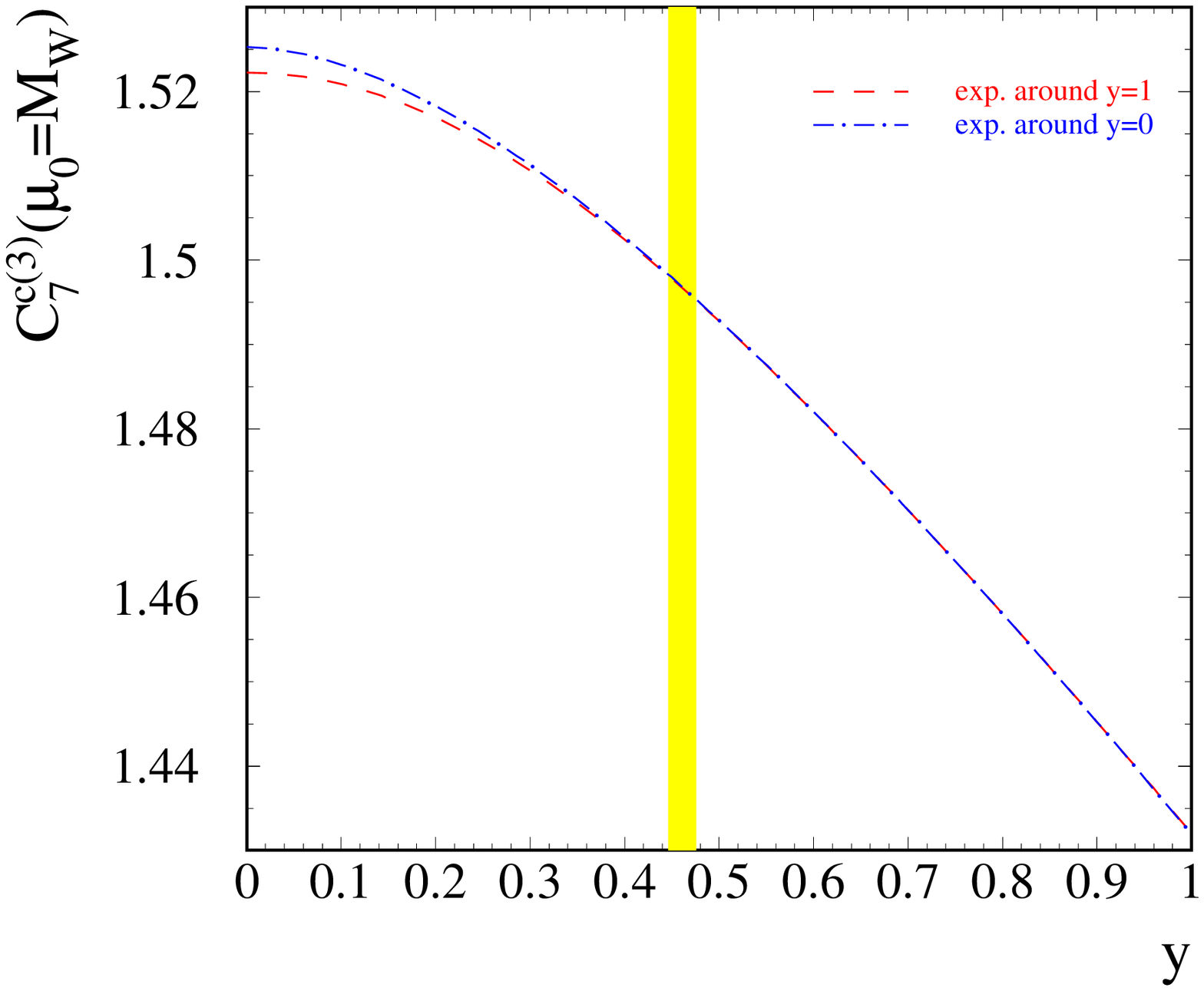}%\hspace{-4cm} \ 
\caption{The coefficients $C_7^{Q(n)}(\mu_0)$ as functions of $y =
M_W/m_t(\mu_0)$. The (blue) dot-dashed lines correspond to their
expansions in $y$ up to $y^8$. The (red) dashed lines describe the
expansions in $(1-y^2)$ up to $(1-y^2)^8$. The (black) solid lines in
the one- and two-loop cases correspond to the known exact
expressions. The (yellow) vertical strips indicate the experimental
range for $y$.} 
\label{fig:c7}
\end{center}
\vspace*{-.5cm}
\end{figure*}

The matching coefficients are obtained from
the requirement that the one-particle-irreducible Green functions in the full
and effective theory are equal. In the latter case only
tree-level diagrams contribute as after expansion in the external momenta
only scale-less integrals appear which are set to zero in  
dimensional regularization~\cite{Gorishnii:1986gn}.
However, the counterterm contributions that arise
from the effective-theory side have to be taken into account.
On the SM side we have to consider of the order of 1000 three-loop diagrams.
One of which is shown in Fig.~\ref{fig:sampleSM}.
Obviously, when the virtual top quark is present in the open fermion
line, we have to deal with three-loop vacuum integrals involving two
mass scales, $m_t$ and $M_W$. Also in the charm-quark sector
such two-scale integrals are
encountered when closed top-quark
loops arise on the virtual gluon lines.
At present, complete three-loop algorithms exist for vacuum integrals
involving only a single mass scale. We have reduced our calculation to
such integrals by performing expansions around the point $m_t=M_W$ and
for $m_t \gg M_W$. In the latter case, the method of asymptotic
expansions of Feynman integrals has been applied \cite{Smi02}
which has been implemented in the {\tt C++} program 
{\tt exp}~\cite{exp}. At the
physical point where $M_W/m_t\approx0.5$, both expansions work
reasonably well.
This can be seen in Fig.~\ref{fig:c7} where the one-, two- and
three-loop results for $C_7$, defined through
\begin{eqnarray}
  C_7^Q &=& 
  C^{Q(0)}_7 
  + \frac{\alpha_s}{4\pi} C^{Q(1)}_7  
%  + \left(\frac{\alpha_s}{4\pi}\right)^2\al^2 C^{Q(2)}_i  
  + \ldots
  \,,
\end{eqnarray}
($Q = c,t$), are shown as functions of $y =
M_W/m_t(\mu_0)$.
(The plots for $C_8$ show an
analog behaviour and can be found in Ref.~\cite{MisSte04}.)

The variable $y$ changes from 0 to 1, i.e. both starting points of our
expansions are present in the figures.  Note the relatively narrow
ranges of the coefficient values on the vertical axes. The large $m_t$
expansions (up to $y^8$) are depicted by the dot-dashed lines, while
the expansions around $m_t = M_W$ (up to $(1-y^2)^8$) are given by the
dashed ones.  In the one- and two-loop cases, solid curves show the
exact results.  The vertical strips mark the experimental values for
$y$.

Comparing the three curves in the one- and two-loop cases, one can
conclude that a combination of 
the two expansions at hand gives a good determination of the studied
coefficients in the whole considered range of $y$.

Although we do not know the exact curves in the three-loop case, the
same pattern seems to repeat. In fact, the charm-sector expansions
perfectly overlap in the physical region. In the top sector, one can
(conservatively) conclude that
\begin{eqnarray}
C_7^{t(3)}(\mu_0=m_t) &=& 12.05 \pm 0.05,\\
C_8^{t(3)}(\mu_0=m_t) &=& -1.2 \pm 0.1,
\end{eqnarray}
which is perfectly accurate for any phenomenological application. Let
us note that a change of $C_7^{t(3)}(\mu_0=m_t)$ from 12 to 13 would
affect the $b \to s \gamma$ decay width by only 0.02\%, while a
similar variation of $C_8^{t(3)}(\mu_0=m_t)$ would cause even a smaller
effect.
More detailed results can be found in Ref.~\cite{MisSte04}.

%%%%%%%%%%%%%%%%%%%%%%%%%%%%%%%%%%%%%%%%%%%%%%%%%%%%%%%%%%%%

\section{\label{sec::ferm}Fermionic corrections to $P_1$, $P_2$, $P_7$
  and $P_8$} 

\begin{figure*}[t]
  \begin{center}
  \begin{tabular}{cccc}
  \includegraphics[bb=60 565 400 700,height=1.5cm]{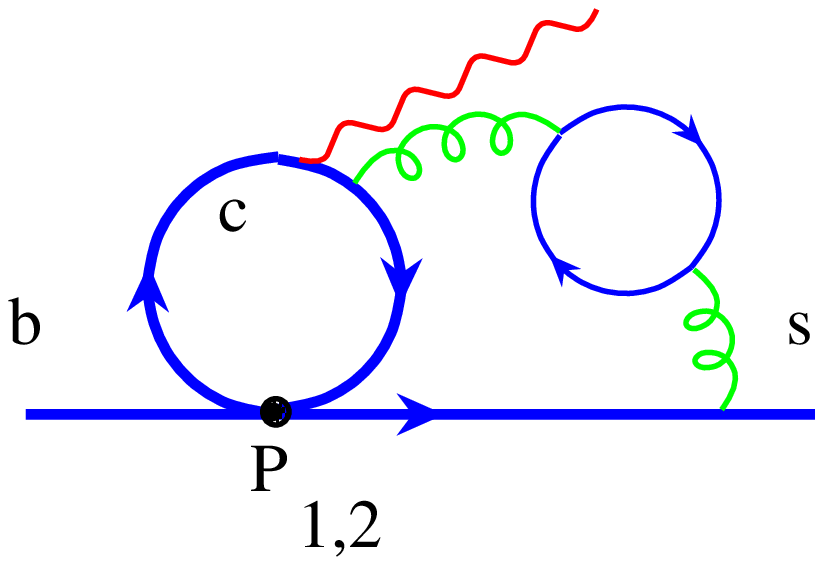}
  &
  \includegraphics[bb=120 565 400 700,height=1.5cm]{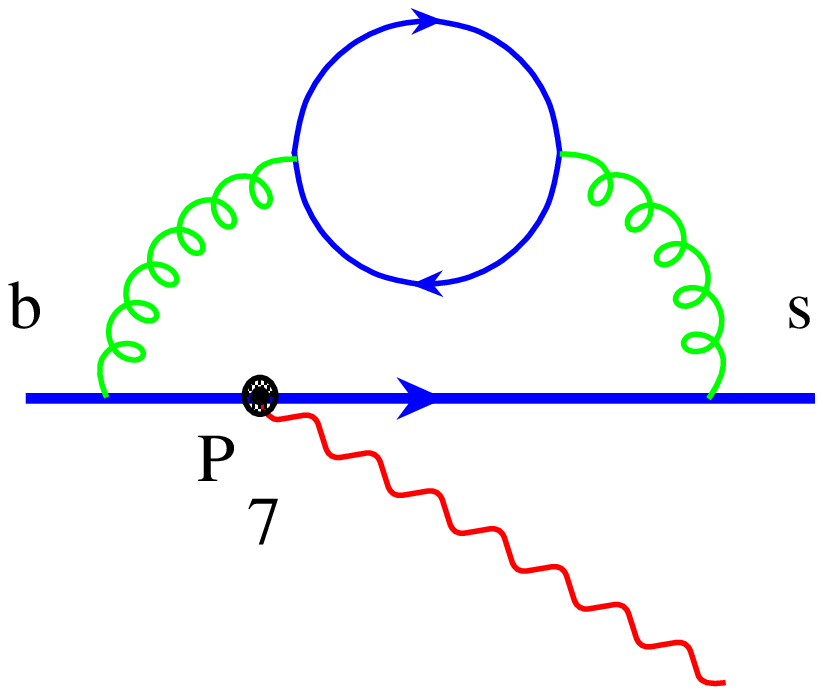}
  &
  \includegraphics[bb=120 565 400 700,height=1.5cm]{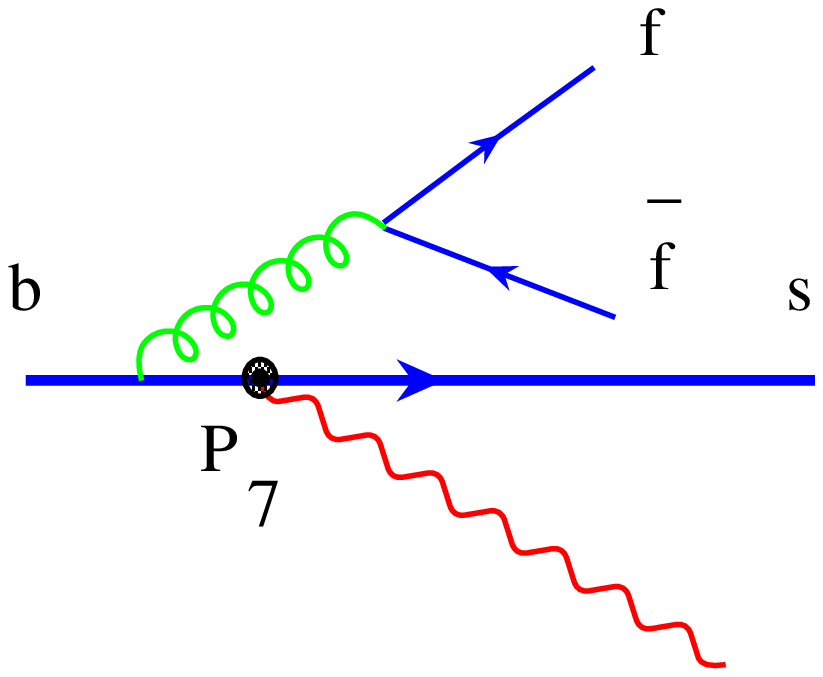}
  &
  \includegraphics[bb=120 565 400 700,height=1.5cm]{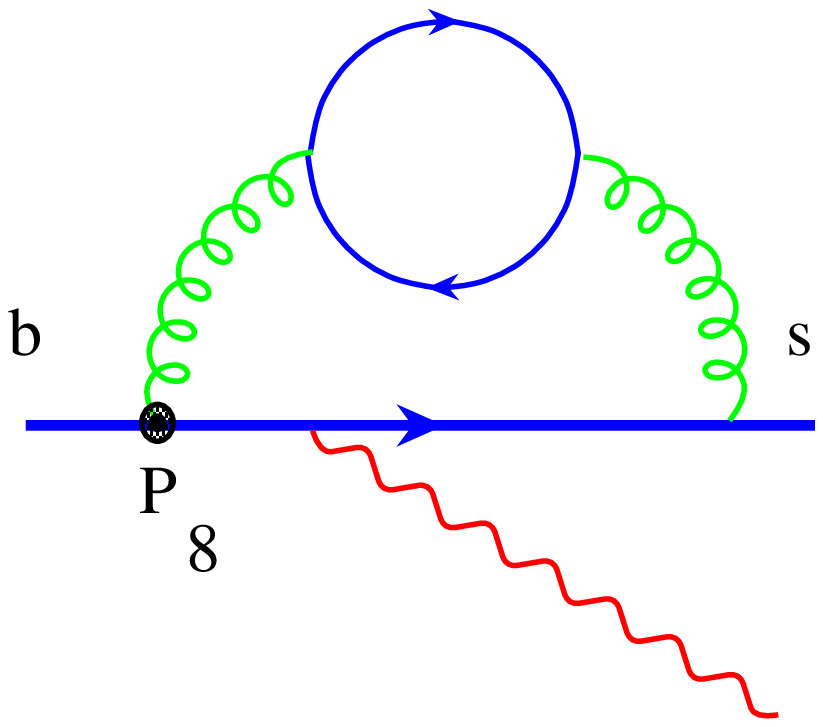}
  \end{tabular}
  \vspace*{.1em}
  \caption{\label{fig::P_i}Sample diagrams contributing to the matrix
    elements of the operators $P_i$ ($i=1,2,7,8$). In the case of $P_7$
    next to the virtual corrections 
    also the real radiation has to be considered. 
    }
  \end{center}
\vspace*{-.5cm}
\end{figure*}

The fermionic NNLO corrections of ${\cal O}(\alpha_s^2 n_f)$
to $P_1$, $P_2$, $P_7$ and $P_8$ (cf. Ref.~\cite{MisSte04} for their
definition) require the
evaluation of two- and three-loop diagrams where a light fermion loop is
inserted into the gluon line. Sample diagrams are shown in
Fig.~\ref{fig::P_i}.

The computation of the virtual corrections proceeds along the same lines as
at NLO~\cite{Greub:1996tg}. The essential difference is
a modified gluon propagator which gets an additional
$\epsilon$-dependent contribution $1/p^{2\epsilon}$
arising from the light fermion loop.

The amplitudes of the contributing diagrams are constructed from
one-loop building blocks describing the  
light-fermion insertion in the gluon propagator and the charm quark
mass-dependent $b\bar{s}g\gamma$-vertex. Afterwards Feynman parameters
are introduced. In the case of $P_7$ and $P_8$ the corresponding
integrations can be performed analytically which is not possible for
$P_1$ and $P_2$. For the latter it is convenient to switch for the
propagator involving the charm and the bottom quark mass
to the Mellin-Barnes representations. This leads to a natural expansion
in $m_c/m_b$ which in practice converges very fast; 
the term of ${\cal O}((m_c^2/m_b^2)^3)$ already leads to a negligible
contribution.

The practical calculation
in the case of $P_7$ is slightly more involved as only the
combination of the virtual corrections with the gluon bremsstrahlung 
and the quark-pair emission process is infrared safe.
We decided to regulate the infrared singularities in the individual
pieces by introducing a finite strange quark mass, $m_s$, and a mass of the
quark in the fermion bubble, $m_f$.
This enables several checks on intermediate results. In particular we
could show that the sum of the virtual
and the gluon bremsstrahlung corrections
are finite in four dimensions in the limit $m_s \to 0$ and for fixed $m_f$.
The remaining dependence on $m_f$ is canceled after including also the 
contribution from the quark-pair emission.
Analytical results for the individual pieces can be found in
Ref.~\cite{Bieri:2003ue}, even for the case where a cut-off in the
photon energy is introduced in the final phase space integration,
which is important for the comparison with the experiment.

\begin{figure*}[t]
  \begin{center}
  \begin{tabular}{cc}
  \includegraphics[bb=30 190 580 590,height=5.0cm]{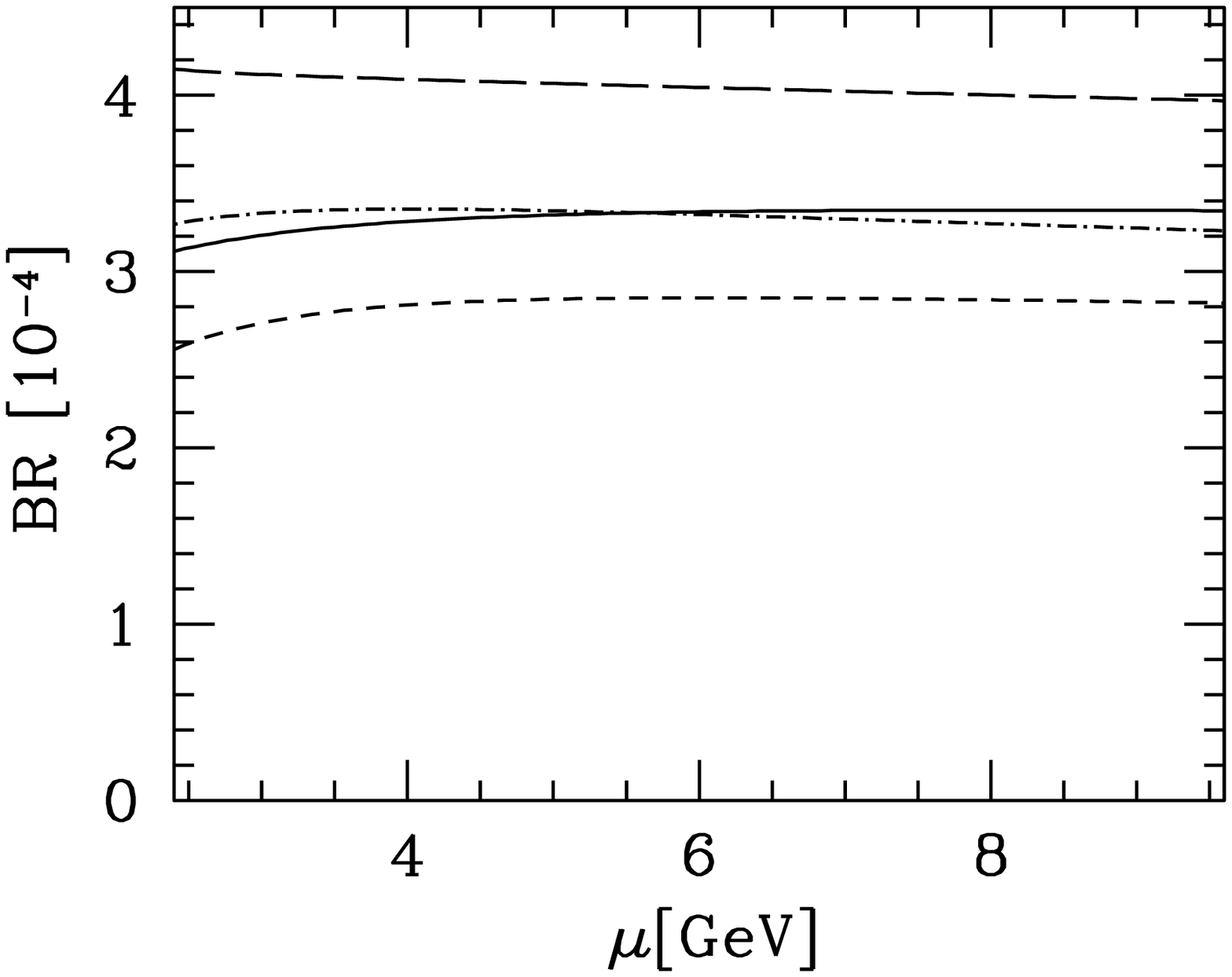}
  &
  \includegraphics[bb=30 190 580 590,height=5.0cm]{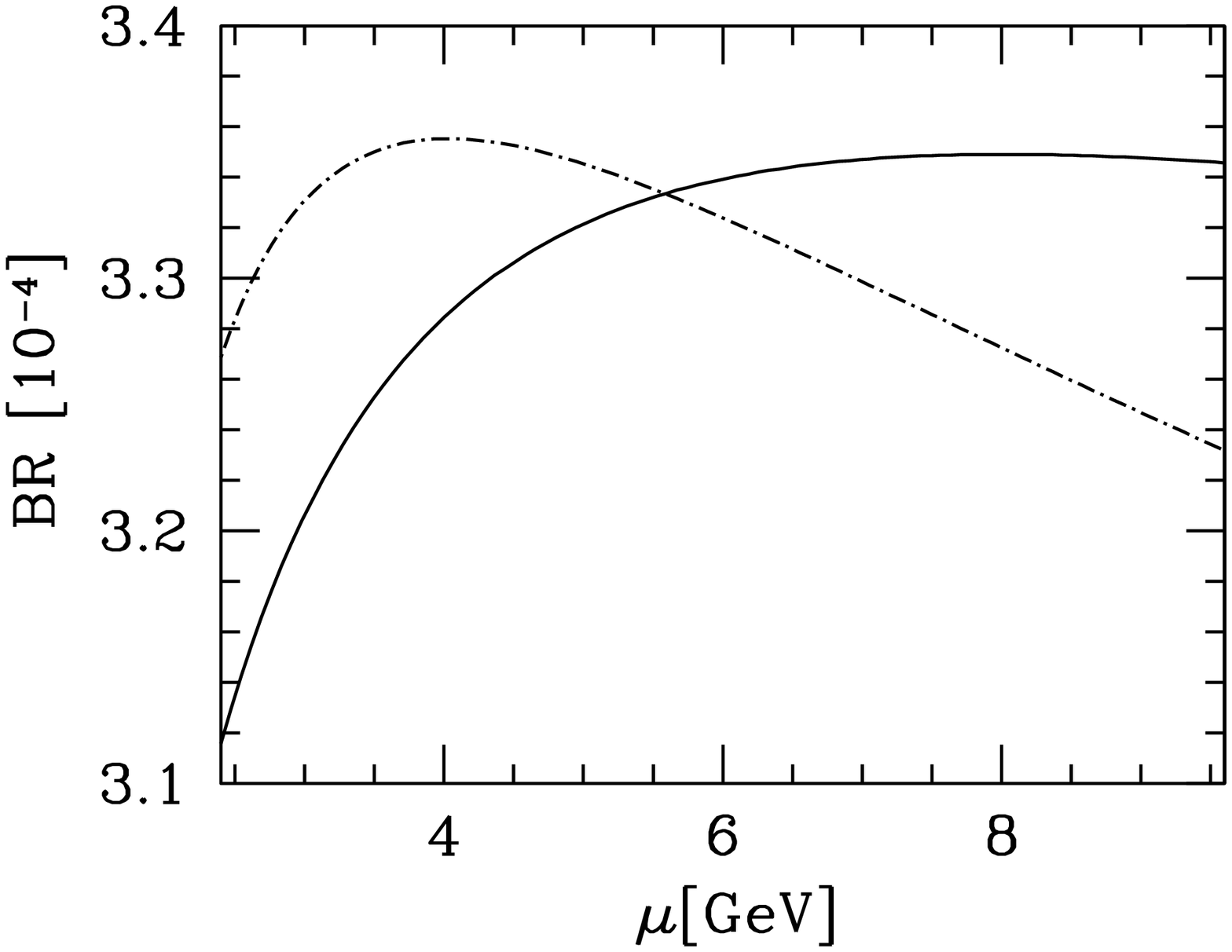}
  \end{tabular}
  \caption{\label{fig:branching}The branching ratio 
    $\mbox{BR}(\bar{B}\to X_s\gamma)$ as a function of the
    renormalization scale $\mu$ where the 
    plot on the right is an enlargement of the one on the left.
    The dash-dotted curve represents the NLO approximation and the 
    solid curve includes the corrections of ${\cal O}(\alpha_s^2 n_f)$.
    For illustration in the left plot
    the latter are also shown for the case where the corrections to 
    $P_{1/2}$ ($P_7$) are set to zero which corresponds
    to short-dashed (long-dashed) curve. A photon energy cut of
  $E_{\rm cut}=m_b/20$ is used, which corresponds to $\delta=0.9$.}
  \end{center}
\vspace*{-.5cm}
\end{figure*}

The numerical effect of the ${\cal O}(\alpha_s^2 n_f)$ corrections is
shown in Fig.~\ref{fig:branching}. The corrections are moderate and
amount to $-3.9\%$ for $\mu=3.0$~GeV and to $+3.4\%$ for $\mu=9.6$~GeV.
Fig.~\ref{fig:branching} illustrates that the $\mu$ dependence of the
${\cal O}(\alpha_s^2 n_f)$ improved prediction for the branching ratio
is somewhat flatter than in the NLO case if we restrict ourselves to 
$\mu\ge4$~GeV. This is a welcome feature of our result, however,
in general we cannot expect to reduce the $\mu$ dependence as the solid curve
only represents a part of the NNLO result.
Indeed, we obtain a stronger $\mu$-dependence in the region below 
$4$~GeV.

%%%%%%%%%%%%%%%%%%%%%%%%%%%%%%%%%%%%%%%%%%%%%%%%%%%%%%%%%%%%

\section{\label{sec::concl}Summary}

In this contribution we reported on the first NNLO calculations to the 
branching ratio $\mbox{BR}(\bar{B}\to X_s\gamma)$. In particular,
the three-loop matching calculations for $C_7$ and $C_8$ have been
described~\cite{MisSte04}. Together with the two-loop calculation for
the other relevant operators calculated several years
ago~\cite{Bobeth:1999mk} this completes the first step towards a
complete NNLO analysis.

As far as the corrections to the operator matrix elements are
concerned up to now only the fermionic corrections to the 
numerically most important operators $P_i$ ($i=1,2,7,8$) have been
evaluated~\cite{Bieri:2003ue}. The remaining
(virtual and real) corrections are not yet available.
The same is true for the evaluation of the anomalous dimension to NNLO
where diagrams up to four loops, from which the pole parts are needed,
have to be computed. 

It should be stressed that only the complete NNLO calculation,
outlined after Eq.~(\ref{eq::lag}), leads to a scheme and scale
independent result and is thus able to reduce the charm quark mass ambiguity.

\section*{Acknowledgements}
I would like to thank K. Bieri, C. Greub and M. Misiak for fruitful
collaborations on the subjects presented in this contribution.
Furthermore, I would like to thank the organizers of Loops and Legs
2004 for the invitation and the nice conference.
This work was supported by HGF Grant No. VH-NH-008.

%%%%%%%%%%%%%%%%%%%%%%%%%%%%%%%%%%%%%%%%%%%%%%%%%%%%%%%%%%%%

\end{document}